\documentclass[preprint,sort&compress,times,fleqn]{elsarticle}

\usepackage{amssymb}
\usepackage{amsmath}
\usepackage{graphicx}
\usepackage{multirow}
\usepackage{subfigure}
\usepackage{float}
\usepackage{graphics}
\usepackage{slashed}
\usepackage{booktabs}
\usepackage{color}
\usepackage[normalem]{ulem}
\usepackage{bm,psfrag}
\usepackage{float}
\usepackage[figuresright]{rotating}

\usepackage{bm}
\usepackage{array}
\usepackage{hyperref}
\usepackage{slashed}
\usepackage{braket}
\usepackage{bigstrut}
\usepackage[nooneline,footnotesize,center]{caption2}

\journal{Nuclear Physics A, published in Nucl. Phys. A 968 (2017) 379-390 }
\begin{document}

\begin{frontmatter}

\title {Effect of sea quarks on single-spin asymmetries $A^{W^{\pm}}_{N}$ in transversely polarized pp collisions at RHIC}

\author[PKU]{Fang Tian}

\author[PKU]{Chang Gong}

\author[PKU,CIC,CHEP]{Bo-Qiang Ma\corref{cor1}}
\ead{mabq@pku.edu.cn}
\cortext[cor1]{Corresponding author at: School of Physics, Peking University, Beijing 100871, China.}

\address[PKU]{School of Physics and State Key Laboratory of Nuclear Physics and
Technology, Peking University, Beijing 100871,
China}
\address[CIC]{Collaborative Innovation Center of Quantum Matter, Beijing, China}
\address[CHEP]{Center for High Energy Physics, Peking University, Beijing 100871, China}

\begin{abstract}
We calculate the single-spin asymmetries $A^{W^{\pm}}_{N}$ of $W^{\pm}$ bosons produced in transversely polarized pp collisions with the valence part of the up (u) and down (d) quark Sivers functions treated by an available parametrization and the light-cone quark spectator-diquark model respectively, while the sea part Sivers functions of u and d quarks treated as parametrization. Comparing our results with those from experimental data at RHIC, we find that the Sivers functions of sea quarks play an important role in the determination of the shapes of $A^{W^{\pm}}_{N}$. It is shown that $A^{W^{-}}_{N}$ is sensitive to u sea Sivers function, while $A^{W^{+}}_{N}$ to d sea Sivers function intuitively. The results show that the contributions of u and d sea Sivers functions are rather sizable and of the same sign, and their signs agree with that of d valence quarks and are opposite to that of u valence quarks.
\end{abstract}

\end{frontmatter}

\section{Introduction}\label{sec:intro}
The spin of the nucleon is an active frontier of high energy physics research. In the past, people believed that all the transverse spin effects should be suppressed at high energies for the incapability to distinguish between the transverse polarization itself and its measurable effects. While in 1970's, the reaction of $\Lambda^{0}$~\cite{Bunce:1976yb} was found to exhibit a strong transversely polarized effect. Ever since then, the transverse single spin asymmetry~(TSSA) has inspired interests both in experimental and theoretical studies. In experimental aspect, the semi-inclusive deep inelastic scattering process~(SIDIS) by the HERMES~\cite{Airapetian:2004tw,Airapetian:2001eg, Airapetian:1999tv,Airapetian:2002mf} and SMC~\cite{Bravar:2000ti} collaborations and the hadronic reactions in pp collision by the STAR collaboration at RHIC~\cite{Bravar:1996ki,Adams:1991rw} have revealed a large transverse single-spin asymmetry with a clear kinematic dependence on the transverse momentum of the hadron as well as on the Feynman variables. Also the related asymmetries are found to be sizable, up to 40$\%$.
In recent experiments, the COMPASS~\cite{Alexakhin:2005iw,Adolph:2016dvl} and STAR~\cite{Adamczyk:2015gyk} collaborations have also revealed a non-ignorable asymmetry in the transversely polarized nucleon process.

There exist three important theoretical explanations for the transverse single-spin asymmetry, including the quark-gluon correlation from higher-twist factorization~\cite{Qiu:1998ia,Qiu:1991pp}, the quark-nucleon spin correlation~\cite{Anselmino:2009pn,Anselmino:2015eoa,Aschenauer:2015ndk} due to the transverse momentum dependent (TMD) parton distribution functions (PDFs) such as the Sivers function~\cite{Sivers:1989cc,Sivers:1990fh} or the transverse momentum dependent (TMD) fragmentation functions (FFs) such as the Collins function~\cite{Collins:1992kk},
and the classical coupling of the orbital and the spin from quarks~\cite{Liang:1989mb}. We adopt the TMDs, especially the Sivers function, which stands for the number of unpolarized quarks in the transversely polarized nucleon, to describe the correlation effect between the spin of the nucleon and the transverse momenta of the inner quarks. Besides, the raise of the Wilson line, which is used to ensure the gauge invariance property of QCD, makes the Sivers function process dependent. Thus a sign change exists between the initial interaction, which describes the interaction between the quark and the remaining nucleon in the Drell-Yan~(D-Y) or Drell-Yan type process~(e.g., $W^{\pm}/Z^{0}$), and  the final interaction, which is related to the interaction between the struck quark and the remaining nucleon in SIDIS~\cite{Brodsky:2002cx,Brodsky:2002rv,Collins:2002kn,Kang:2009bp,Boer:1997nt,Aschenauer:2015ndk}.
Due to the rich experimental data in SIDIS process, the analyses and extractions of Sivers functions with large error bars have been performed for both the sea and the valence parts~\cite{Anselmino:2016uie,Anselmino:2004nk,Anselmino:2008sga,Anselmino:2002pd,Sivers:1989cc,Sivers:1990fh,Vogelsang:2005cs,Collins:2005ie,Anselmino:2005an,Anselmino:2005ea,Anselmino:2005nn,Anselmino:2013rya,Afanasev:2007ii}.
Also there are some model calculations of Sivers distribution functions~\cite{Kumar:2015coa,Lu:2004au,Bacchetta:2003rz,Yuan:2003wk,Brodsky:2002cx,Brodsky:2002rv,Bacchetta:2008af} for both $u$ and $d$ quarks. Though these studies adopt different models, they get similarly results, especially in Refs.~\cite{Lu:2004au,Yuan:2003wk,Bacchetta:2008af}, where both of  $u$ and $d$ Sivers functions are sizable. The size of $d$ Sivers functions is 6-7 times smaller than that of $u$ Sivers functions.
In addition, researchers considered the influence due to the evolution of TMDs on TSSA~\cite{Echevarria:2014xaa,Aidala:2014hva,Collins:2014loa,Sun:2013hua}, and
the gluon Sivers function have been also studied~\cite{Anselmino:2006yq,Lu:2016vqu,Burkardt:2004ur,DAlesio:2015fwo,Burkardt:2004ur}.

The TSSA have been observed in transversely polarized $p^{\uparrow}p$ collision process at RHIC~\cite{Adamczyk:2015gyk}, and there have been some discussions on these experimental results~\cite{Huang:2015vpy, Anselmino:2016uie}. There are still some discrepancies between the theoretical calculations and the experimental data. Just as the similar situation in the single spin asymmetries in longitudinally polarized $\vec{p}p$ collisions~\cite{Tian:2017xul}, we suspect that the large sea Sivers functions of $\bar{u}$ and $\bar{d}$ quarks should be needed in $W^{\pm}$ production at RHIC. So we adopt two different procedures to extract the sea Sivers functions, namely, we treat the valence part of $u$ and $d$ quark Sivers functions with an available parametrization and with the light-cone quark spectator-diquark model respectively, while the sea part Sivers functions of $u$ and $d$ quarks are fitted by parametrization. It is interesting to find that the two procedures produce similar results about the sea Sivers functions of both $u$ and $d$ quarks.

In this paper, we investigate the contribution of sea quark Sivers functions to $A^{W^{\pm}}_{N}$ with the valence part modeled by the parameterization and the quark spectator-diquark model. Sec.~\ref{sec:ww1} presents the definitions and parameterizations of Sivers functions. Sec.~\ref{sec:ww2} presents the necessary formulae of the spin asymmetry, as well as the extractions of sea quark Sivers functions from the corresponding single spin asymmetries in $W^{\pm}$ production processes. For convenience, we use $\Delta^N q$ as a short notation for the Sivers functions of quarks.
We find that the shape of $A^{W^{-}}_{N}$ is sensitive to $\Delta^N \bar u$, while the shape of $A^{W^{+}}_{N}$ is sensitive to $\Delta^N \bar d$ intuitively. Both $\Delta^N \bar u$ and $\Delta^N \bar d$ are rather sizable and of the same sign with the valence $\Delta^N d$ and opposite to valence $\Delta^N u$ for better description of experimental data. Numerical results and discussions are presented. A summary is given in the final section.

\section{Sivers functions}\label{sec:ww1}
As we know, the Sivers distribution function plays an important role in explaining the transverse single spin asymmetries in hadronic processes. It describes the correlation effect between the spin of nucleons and the transverse momenta of quarks.
The general expression for the number density
of quarks with flavor $q$ and transverse momentum $\bf{k_{\perp}}$, inside a proton with spin ${\bf{S}}$ and
three-momentum ${\bf{p}}$, is:
\begin{eqnarray}
\Delta^Nf_{q/p^{\uparrow}}(x,{\bf{k_{\perp}}}) &=&
\hat f_{q/p^{\uparrow}}(x,{\bf{k_{\perp}}})-\hat f_{q/p^{\downarrow}}(x, {\bf{k_{\perp}}}) =
\hat f_{q/p^{\uparrow}}(x, {\bf{k_{\perp}}})-\hat f_{q/p^{\uparrow}}(x, -{\bf{k_{\perp}}}),\,\nonumber \\
\hat f_{q/p^{\uparrow}}(x, {\bf{k_{\perp}}})&=& \hat f_{q/p}(x, k_{\perp})
+ \frac{1}{2} \Delta^N f_{q/p^{\uparrow}}(x, k_{\perp})  \hat{\bf{S}} \cdot
\hat{\bf{p}} \times \hat{\bf{k}}_{\perp}\\
&=&\hat f_{q/p}(x, k_{\perp})
+\frac{k_{\perp}}{m_p}f_{1T}^{\perp q}(x, k_{\perp})  \hat{\bf{S}} \cdot
\hat{\bf{p}} \times \hat{\bf{k}}_{\perp},
\label{eq:01}
\end{eqnarray}
here $\Delta^Nf_{q/p^\uparrow}(x,k_{\perp})$ is the so called Sivers distribution function~\cite{Anselmino:2002pd,Sivers:1989cc,Sivers:1990fh}.
For convenience in description, we adopt the sign convention of $f_{1T}^{\perp q}$ along with Refs.~\cite{Lu:2004au,Martin:2017yms}, where the Sivers function is positive for $u$ valence quarks and negative for $d$ valence quarks, in analogy with the signs of $u$ and $d$ valence quark helicity distributions.

\subsection{Parameterizations}
Sivers functions can be extracted or parameterized from SIDIS data~\cite{Sivers:1989cc,Sivers:1990fh,Vogelsang:2005cs,Collins:2005ie,Anselmino:2008sga,Anselmino:2016uie,Martin:2017yms}.
To estimate the Sivers functions from data, we introduce the following two different parameterizations, e.g., a point-by-point form in Refs.~\cite{Martin:2014wua,Martin:2017yms}, and a general form in Ref.~\cite{Anselmino:2016uie}.

The point-by-point method, which is parameter-free, relies on the simple assumption of the Gaussian behavior in the transverse momenta and the experimental data in SIDIS. The Sivers valence and the isotriplet $\bar{u}-\bar{d}$ component of the Sivers sea can be extracted at fixed-point by solving the equations of the spin asymmetry. Through this method, $u_v$ Sivers function is positive~(black solid circles) and $d_v$ Sivers function is negative~(black open circles), as shown in Fig.~\ref{fig:a02}, and the results are consistent with following parameterization and model calculations.
The value of the Sivers sea $\bar{u}-\bar{d}$ is compatible with zero.


The general parameterized form of Sivers functions~\cite{Anselmino:2016uie} is usually based on the following formulae:
\begin{eqnarray}
\begin{aligned}
\Delta^{N} f_{q/p^{\uparrow}}(x,{k}_{\perp})=2\mathcal{N}_{q}(x)h(k_{\perp})f_{q/p}(x,k_{\perp});\\
\mathcal{N}_{q}(x)=N_{q}x^{\alpha_{q}}(1-x)^{\beta_{q}}\frac{(\alpha_{q}+\beta_{q})^{(\alpha_{q}+\beta_{q})}}{{\alpha_{q}}^{\alpha_{q}}{\beta_{q}}^{\beta_{q}}};\\
h(k_{\perp})=\sqrt{2e}\frac{k_{\perp}}{M_1}\exp{(-k^{2}_{\perp}/M^{2}_{1})},
\label{eq:02}
\end{aligned}
\end{eqnarray}
where $N_q$, $\alpha_q$, $\beta_q$ and $M_1$ (GeV/$c$) are free parameters
to be determined by fitting the experimental data. Since $h(k_{\perp}) \le 1$ for any $k_{\perp}$ and $|{\cal N}_q(x)| \le 1$ for any $x$ (notice that we allow the constant parameter $N_q$ to vary only inside the range $[-1,1]$), the
positivity bound for the Sivers function,
\begin{eqnarray}
\begin{aligned}
\frac{|\Delta^N f_ {q/p^{\uparrow}}(x,k_{\perp})|}{2 f_ {q/p} (x,k_{\perp})} \le 1,
\label{eq:pos}
\end{aligned}
\end{eqnarray}
is automatically fulfilled.
The unpolarized TMDs are expressed as:
\begin{eqnarray}
\begin{aligned}
f_{q/p}(x,k_\perp) = f_q(x) \, \frac{1}{\pi \langle k_\perp^2\rangle} \,
e^{-k_\perp^2/\langle k_\perp^2\rangle}.
\label{eq:03}
\end{aligned}
\end{eqnarray}
The relevant parameters of valence quarks are from Ref.~\cite{Anselmino:2016uie}:
$N_{u_v} = 0.18 \pm{0.01}{\pm0.04} $, $N_{d_v} = -0.52 \pm{0.08}{\pm0.20}$, $\alpha_{u_v} = 1.0 \pm{0.3}{\pm0.6}$, $\alpha_{d_v} = 1.9 \pm{0.5}{\pm1.5}$, $\beta_{u_v}= 6.6\pm{2.0}{\pm5.2}$, $\beta_{d_v}= 10.0\pm{4.0}{\pm11.0}$, $M_1^2 = 0.8 {\pm{0.20}}{\pm{0.9}}$ (GeV/$c)^2$, and $\langle k_{\perp}^2\rangle   = 0.57\pm{0.08}$.
During the later fitting period, we neglect the TMD evolution and just consider the collinear evolution of unpolarized distribution functions. Besides, we adopt CTEQ parametrization (CT14LO)~\cite{Dulat:2015mca} as an example for the input unpolarized PDFs.

\subsection{Model calculations}
In Ref.~\cite{Lu:2004au}, the valence $u$ and $d$ Sivers functions are predicted based on the SU(6) quark spectator-diquark model (denoted as qD model) by considering both scalar and vector diquarks. The basic formulae for the valence Sivers functions are:
\begin{equation}
f^{\perp
a}_{1T}(x,\mathbf{k}_\perp^2)=-\frac{M\mathcal{P}_y^a}{\mathbf{k}_\perp^1}f_1^a(x,\mathbf{k}_\perp^2)\label{pya}.
\end{equation}

\begin{equation}
\mathcal{P}_y^{u/d}=\frac{\textmd{Im}\Big{(}\mathcal{T}_{\mathrm{int}}^{u/d}\Big{)}}{\mathcal{M}^{u/d}}.\label{eq:py}
\end{equation}

\begin{eqnarray}
\mathcal{T}_{\mathrm{int}}^u&=&-\frac{1}{2}(M+\frac{m}{x})\frac{\mathbf{k}_\perp^1}{x}
\frac{e_1e_2}{4\pi}\frac{1}{\Lambda_s(\mathbf{k}^2_\perp)\mathbf{k}_\perp^2}
\textmd{ln}\frac{\Lambda_s(\mathbf{k}^2_\perp)}{\Lambda_s(0)}\nonumber\\
&&-\Big{[}\frac{2}{9}(M+\frac{\lambda_V}{1-x})\frac{\mathbf{k}_\perp^1}{1-x}
+\frac{1}{9}(M+2\frac{\lambda_V}{1-x}-\frac{m}{x})\nonumber\\
&&\cdot\frac{1-x}
{(1+x)}\frac{\mathbf{k}_\perp^1}{x}\Big{]}\frac{e_1e_2}{8\pi}\frac{1}{\Lambda_v(\mathbf{k}^2_\perp)\mathbf{k}_\perp^2}
\textmd{ln}\frac{\Lambda_v(\mathbf{k}^2_\perp)}{\Lambda_v(0)},\\
\mathcal{T}_{\mathrm{\mathrm{int}}}^d&=&-\frac{e_1e_2}{4\pi}\frac{1}{\Lambda_v(\mathbf{k}^2_\perp)
\mathbf{k}^2_\perp}\textmd{ln}\frac{\Lambda_v(\mathbf{k}^2_\perp)}{\Lambda_v(0)}
\Big{[}\frac{2}{9}(M+\frac{\lambda_V}{1-x})\frac{\mathbf{k}_\perp^1}{1-x}\nonumber\\&
&+\frac{1}{9}(M+2\frac{\lambda_V}{1-x}-\frac{m}{x})\frac{1-x}
{(1+x)}\frac{\mathbf{k}_\perp^1}{x}\Big{]},\\
\mathcal{M}^u&=&\frac{1}{2}\Big{[}(M+\frac{m}{x})^2+\frac{\mathbf{k}_\perp^2}
{x^2}\Big{]}h_s^2+\frac{1}{9}\Big{[}(M+\frac{\lambda_V}{1-x})^2
+\frac{\mathbf{k}_\perp^2}{(1-x)^2} \nonumber\\&
&+\frac{\mathbf{k}_\perp^2}{x^2(1-x)^2}
+\frac{(1-x)^2}{(1+x)^2}\frac{\mathbf{k}_\perp^2}{2x^2}+(\frac{1}{2}M+\frac{\lambda_V}{1-x}-\frac{m}{2x})^2\bigg{]}h_v^2,\\
\mathcal{M}^d&=&\frac{2}{9}\Big{[}(M+\frac{\lambda_V}{1-x})^2+\frac{\mathbf{k}_\perp^2}
{(1-x)^2}+\frac{\mathbf{k}_\perp^2}{x^2(1-x)^2}
+\frac{(1-x)^2}{(1+x)^2}\frac{\mathbf{k}_\perp^2}{2x^2}\nonumber\\&
&+(\frac{1}{2}M+\frac{\lambda_V}{1-x}-\frac{m}{2x})^2\bigg{]}h_v^2,
\end{eqnarray}
in which
\begin{eqnarray}
\Lambda_{s/v}(\mathbf{k}^2_\perp)=\mathbf{k}^2_\perp+x(1-x)(-M^2+\frac{m^2}{x}+\frac{\lambda^2_{S/V}}{1-x}),\nonumber\\
h_{s/v}=\frac{1}{\mathbf{k}^{2}_\perp+x(1-x)(-M^2+\frac{m^2}{x}+\frac{\lambda^2_{S/V}}{1-x})}.
\end{eqnarray}
The related parameters are:
\begin{eqnarray}
\begin{aligned}
\lambda_S=0.6\;{\rm GeV}, \quad
\lambda_V=0.9\;{\rm GeV}, \quad\\
m=0.36\;{\rm GeV}, \quad
M=0.94\;{\rm GeV} \>,
\label{eq:05}
\end{aligned}
\end{eqnarray}
here $\lambda_S$, $\lambda_V$, $m$ and $M$ are the masses of the scalar diquark, the vector diquark, the quark and the proton respectively.
We adopt the value of $\langle k_\perp^2\rangle$
fixed in SIDIS~\cite{Anselmino:2002pd,Anselmino:2008sga,Anselmino:2016uie}:
$\langle k_{\perp}^2\rangle   = 0.57  \;({\rm GeV}/c)^2$.

\begin{figure}[H]
	\begin{center}
	\includegraphics[width=0.40\textwidth]{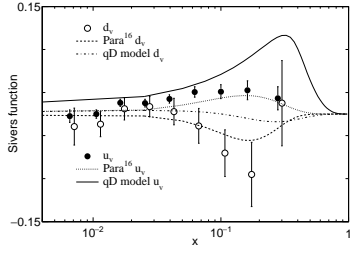}
	\end{center}
	\vspace{-0.5cm}
	\caption{\label{fig:a02} $xf^{\perp(1)}_{1T}$ at $Q=2$ GeV. The scripts $\mathrm{para^{16}}$ are the valence Sivers functions of $u$ and $d$ from~\cite{Anselmino:2016uie}; $\mathrm{qD~model}$ are the valence Sivers functions of $u$ and $d$ from~\cite{Lu:2004au}, and the open~(solid) circles denote the extracted results of valence Sivers functions from~\cite{Martin:2017yms}.}
\end{figure}

The related comparisons of valence quark Sivers functions between the parameterizations~\cite{Anselmino:2016uie} and the model calculations~\cite{Lu:2004au} are
presented in Fig.~\ref{fig:a02}.
From Fig.~\ref{fig:a02}, we notice that the results of the parameterization~\cite{Anselmino:2016uie} can be comparable
with those from the extractions~\cite{Martin:2014wua,Martin:2017yms}, while the results of the model calculations are different from them. But in all of the three cases, the Sivers functions are of positive sign for $u$ valence quarks and negative sign for $d$ valence quarks. 

\section{Results}\label{sec:ww2}
\subsection{$A_{N}$ in $W^{\pm}$ production process}
The $W^{\pm}$ production process has a clean final state without any fragmentation process, therefore the $W^{\pm}$ production process can be used as a key tool to study the TSSA.
The TSSA for $W^{\pm}$ production process is given by:
\begin{eqnarray}
\begin{aligned}
A_N = \frac{d\sigma^\uparrow - d\sigma^\downarrow}
{d\sigma^\uparrow + d\sigma^\downarrow}=\frac{d\sigma^\uparrow - d\sigma^\downarrow}
{2d\sigma},
\end{aligned}
\label{eq:01}
\end{eqnarray}
here the subscripts $\uparrow$ $(\downarrow)$ mean the transversely polarized direction of the initial nucleon.

The factorization theorem of cross sections based on the TMDs in $W^{\pm}$ production process $p^{\uparrow}p\rightarrow W^{\pm} X\rightarrow \ell^{\pm}\nu X$ is~\cite{Anselmino:2009pn}:
\begin{eqnarray}
\begin{aligned}
d\sigma = \sum_{ab} \hat f_{a/p}(x_a,{\bf{k}}_{\perp a}) \,\otimes
\hat f_{b/p}(x_b,{\bf{k}}_{\perp b}) \,\otimes
d\hat\sigma^{ab \to\ell^{\pm}\nu},
\end{aligned}
\label{eq:06}
\end{eqnarray}
here $\hat f_{a/p}(x_a,{\bf{k}}_{\perp a})$ $(\hat f_{b/p}(x_b,{\bf{k}}_{\perp b}))$ mean the parton distribution functions with longitudinal momentum fractions $x_a(x_b)$ and transverse momenta $\bf{k}_{\perp a}(\bf{k}_{\perp b})$, i.e., they are unpolarized TMDs. $d\hat\sigma^{ab \to\ell^{\pm}\nu}$ is the cross section for the parton process $ab \rightarrow \ell^{\pm}\nu$.

So the unpolarized cross section is:
\begin{eqnarray}
\begin{aligned}
d\sigma = \sum_{ab} \int \left[ dx_a \, d^2{\bf{k}}_{\perp a}
\, dx_b \, d^2{\bf{k}}_{\perp b} \right] \, \hat f_{a/p}(x_a,{\bf{k}}_{\perp a}) \,
\hat f_{b/p}(x_b,{\bf{k}}_{\perp b}) \,
d\hat\sigma^{ab \to \ell^{\pm}\nu}.
\end{aligned}
\label{eq:07}
\end{eqnarray}
Also the transverse spin dependent cross section is:
\begin{eqnarray}
\begin{aligned}
d\sigma^\uparrow - d\sigma^\downarrow =
\sum_{ab} \int \left[ dx_a \, d^2{\bf{k}}_{\perp a}
\, dx_b \, d^2{\bf{k}}_{\perp b} \right] \,
\Delta^Nf_{a/p^\uparrow}(x_a,{\bf{k}}_{\perp a}) \,
\hat f_{b/p}(x_b,{\bf{k}}_{\perp b}) \,
d\hat\sigma^{ab \to \ell^{\pm}\nu} .
\end{aligned}
\label{eq:08}
\end{eqnarray}
So the expression of TSSA is:
\begin{eqnarray}
A^{W^{\pm}}_{N}&=&\frac{d\sigma^{\uparrow}-d\sigma^{\downarrow}}{2d\sigma}\,\nonumber\\
&=&\frac{\sum_{ab}\,\int\left[dx_adx_bd^2{\bf{k}}_{\perp a}d^2{\bf{k}}_{\perp b}\right]\Delta^{N} f_{a/p^{\uparrow}}(x_a,{\bf{k}}_{\perp a})\,\times
\hat f_{b/p}(x_b,{\bf{k}}_{\perp b})\,\times
d\hat\sigma^{ab \to \ell^{\pm}\nu}\nonumber}{2\sum_{ab}\,\int\left[dx_adx_bd^2{\bf{k}}_{\perp a}d^2{\bf{k}}_{\perp b}\right]\hat f_{a/p}(x_a,{\bf{k}}_{\perp a})\,\times
\hat f_{b/p}(x_b,{\bf{k}}_{\perp b})\,\times
d\hat\sigma^{ab \to \ell^{\pm}\nu}}\, ,\\
\label{eq:0445}
\end{eqnarray}
here $x_a =
M \, e^y/\sqrt s$, $x_b = M \, e^{-y}/\sqrt s$, and $y$ is the rapidity of the $W$ boson.

After above discussions, the TSSA is directly related to the Sivers distribution functions.
During our calculations, we adopt the pp center-of-mass frame. The momentum of the polarized incoming proton is along the $z$ axis, while the spin of the proton is along the positive $y$-axis. 
Other variables are from the experiment at RHIC: the mass of center-of-mass frame $\sqrt{s}=500$~GeV, and the momentum of the produced $W$ boson $0.5\leq q_T\leq 10$~GeV.

Besides, the cross section $\hat{\sigma}_0$ of the parton process $ab\rightarrow W^{\pm}\rightarrow \ell^{\pm}\nu$ is
\begin{eqnarray}
\begin{aligned}
\hat{\sigma}_{0}=|V_{ab}|^2\frac{\sqrt{2}\pi G_F M^{2}_{W}}{3s},
\label{eq:09}
\end{aligned}
\end{eqnarray}
here $|V_{ab}|$ is the weak interaction quark mixing (CKM) matrix elements, $G_F$ is the Fermi weak coupling constant, and $M_W$ is the mass of $W$ boson.

But the direct Sivers functions for the Drell-Yan and Drell-Yan type $W^{\pm}/Z^{0}$ production processes are still unknown. Upon the gauge invariance of QCD, there exists a sign change between SIDIS and Drell-Yan~(including $W^{\pm}/Z^{0}$ production) process~\cite{Collins:2002kn,Kang:2009bp,Anselmino:2009st,Brodsky:2002cx,Boer:2002ju,Brodsky:2002rv,Anselmino:2016uie}.
Based on this theorem, we can simply use the Sivers function of SIDIS with a sign change to study $W^{\pm}/Z^{0}$ production processes.
For the sake of consistency, our convention of the Sivers functions keep unchanged as in the SIDIS situation in our following discussions and statements.

To account for the evolution effects of $A^{W^{\pm}}_{N}$, we need to consider both the evolutions of unpolarized TMDs and the Sivers functions. The CTEQ parametrization (CT14LO), as an input of unpolarized PDFs, takes the QCD evolution effects of unpolarized PDFs into account. During our fittings of the Sivers functions and unpolarized TMDs, we just consider the collinear evolution of unpolarized PDFs. This is not strict because the evolutions of Sivers function and TMDs are different from that of the unpolarized PDFs. As the evolution effects in the numerator and the denominator of the asymmetry $A^{W^{\pm}}_{N}$ may cancel each other in some ways, our results can be considered as a reasonable reflection of the evolution effects.
Besides, as noted in Refs.~\cite{Aybat:2011zv,Aybat:2011ge,Echevarria:2014xaa,Anselmino:2016uie}, the evolutions of unpolarized TMDs and the Sivers functions may produce a suppression of the asymmetries with increasing $Q^2$. So the evolution effect can not explain the large asymmetries observed in experiments. In this sense, our attempt to understand the physical mechanism for the large asymmetries can not be eliminated by the evolution effect.



 \subsection{Numerical calculations}
 We adopt the forms of sea quark Sivers functions as described by Eq.~(\ref{eq:02}) and Eq.~(\ref{eq:03}) in our numerical calculations with the valence parts fixed by parameterizations in Ref.~\cite{Anselmino:2016uie} and model calculations in Ref.~\cite{Lu:2004au} respectively. The parameters for two different modes of sea and valence quark Sivers functions are given in Tab.~\ref{table1}. We present our numerical results as follows.

 In Table~\ref{table1}, $N_{\bar{u}/\bar{d}}$, $\alpha_{\bar{u}/\bar{d}}$,  $\beta_{\bar{u}/\bar{d}}$ and $M_{1}$ are obtained by fitting experimental data of transversely polarized single-spin asymmetries in $W^{\pm}$ boson production at RHIC~\cite{Adamczyk:2015gyk}. From Table~\ref{table1}, we know that different modes indicate different cases of valence parts. For example, $\mathrm{Mode=1}$ and $\mathrm{Mode=2}$ correspond to the valence parts from the parameterization in Ref.~\cite{Anselmino:2016uie} and model calculations in Ref.~\cite{Lu:2004au} respectively.

 \begin{table}[H]\scriptsize
 	\begin{center}
 		\begin{tabular}{c|c|c|c|c|c|c|c|c|c|c}
 			\hline \hline
 			\multirow{3}{*}{$\Delta^{N} q_V$} & \multirow{3}{*}{$\mathrm{Mode}$}& \multirow{3}{*}{$\langle k_{\perp}^2\rangle$} & \multirow{3}{*}{Data} & \multicolumn{7}{c}{Parameter} \\
 			\cline{5-11}
 			&& && $N_{\bar{u}}$ &$N_{\bar{d}}$ &$\alpha_{\bar{u}}$& $\alpha_{\bar{d}}$& $\beta_{\bar{u}}$& $\beta_{\bar{d}}$ &$M_1$\\
 			\hline
 			\multirow{1}{*}{$\mathrm{Para^{16}}$}
 			&1&\multirow{1}{*}{0.57}&$\mathrm{{W^{\pm}}}$& -1.0 & -1.0& 4.559& 3.265 & 14.97 & 14.903&$\sqrt{0.8}$\\					
 			\hline
 			\multirow{1}{*}{$\mathrm{qD~~model}$} &2&\multirow{1}{*}{0.57}&$\mathrm{{W^{\pm}}}$& -1.0& -1.0&3.338 &3.293 & 11.370 &14.185&1.241\\
 			\hline
 			\hline \hline
 		\end{tabular}
 	\end{center}
 	\caption{\label{table1} Parameters of sea Sivers functions.}
 \end{table}  		
Comparing the results in our modes, we can see that different procedures can produce different distributions of $\Delta^{N}\bar{q}(x,k_{\perp})$. But the basic signs of sea quark Sivers functions are the same. What is more, the results as shown in our modes are consistent with the parameterizations in Ref.~\cite{Anselmino:2016uie}, while the values seem to be larger.

\begin{figure}[H]
	\begin{center}		
\subfigure[$A^{N}_{W^{-}}$.]{\includegraphics[width=0.40\textwidth]{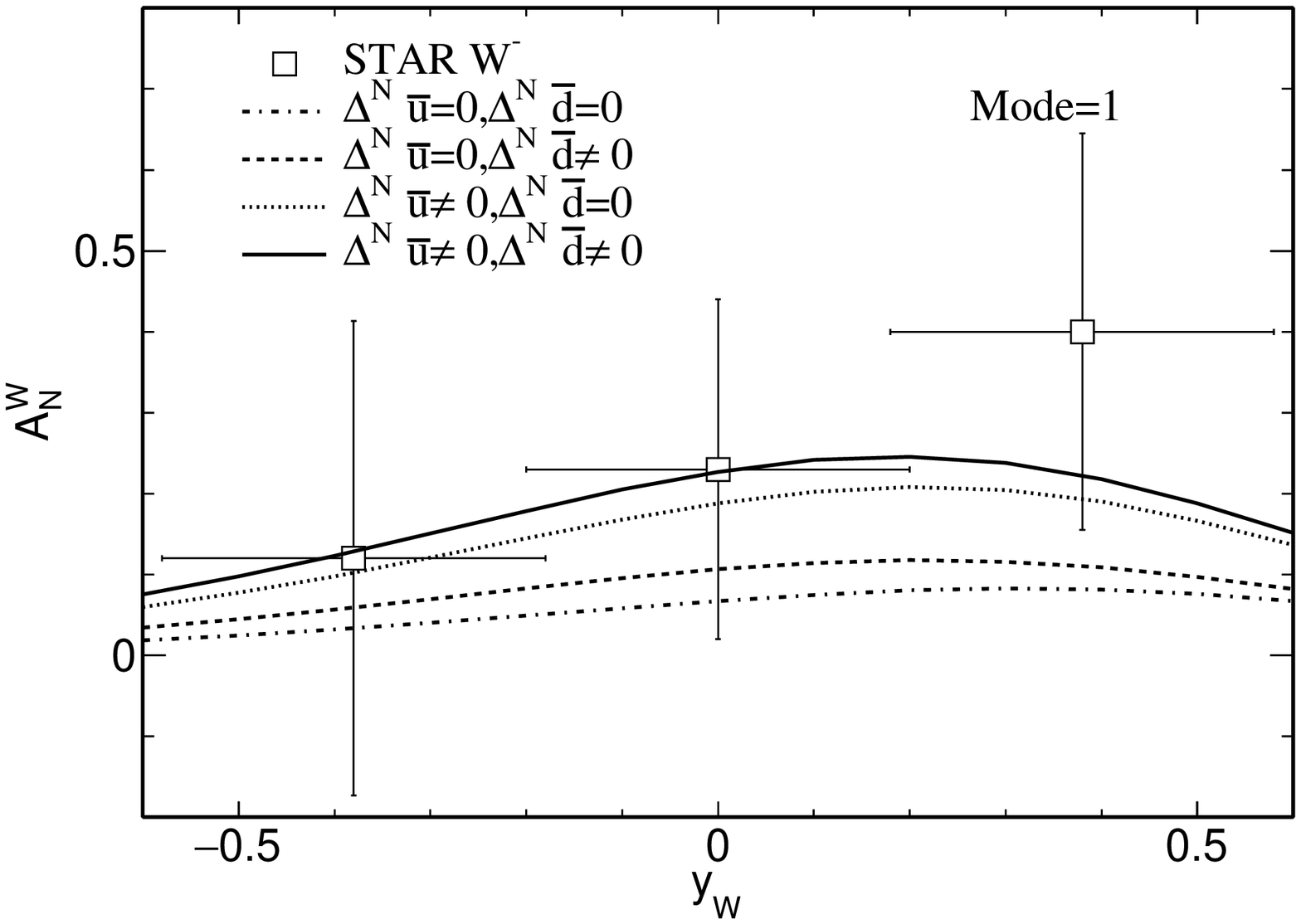}}		\subfigure[$A^{N}_{W^{+}}$.]{\includegraphics[width=0.40\textwidth]{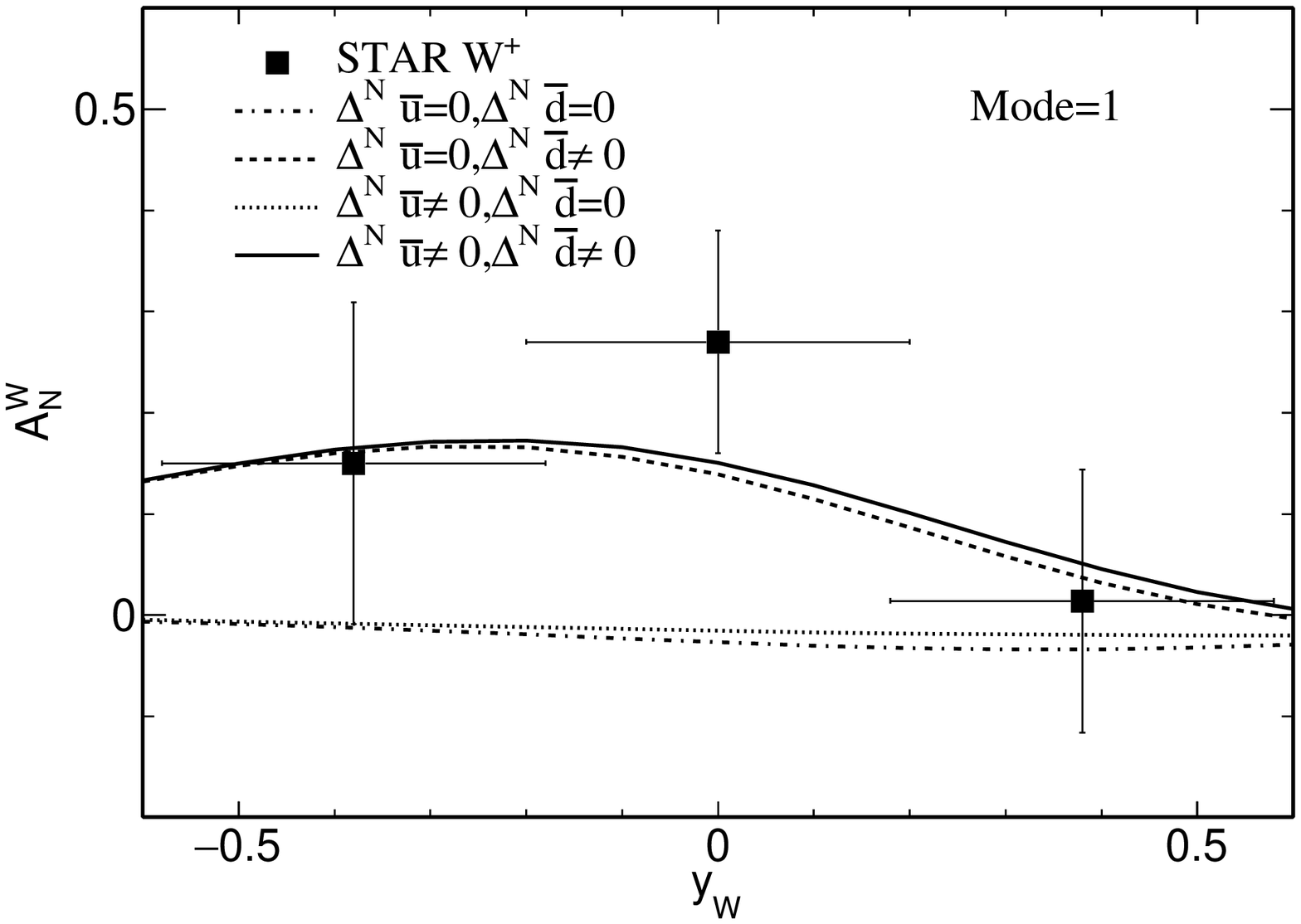}}
	\end{center}
	\vspace{-0.5cm}
	\caption{\label{fig:21} $A^{N}_{W^{\pm}}$ at $Q=M_W$ GeV for $\mathrm{Mode=1}$.}
\end{figure}

\begin{figure}[H]
	\begin{center}		\subfigure[$A^{N}_{W^{-}}$.]{\includegraphics[width=0.40\textwidth]{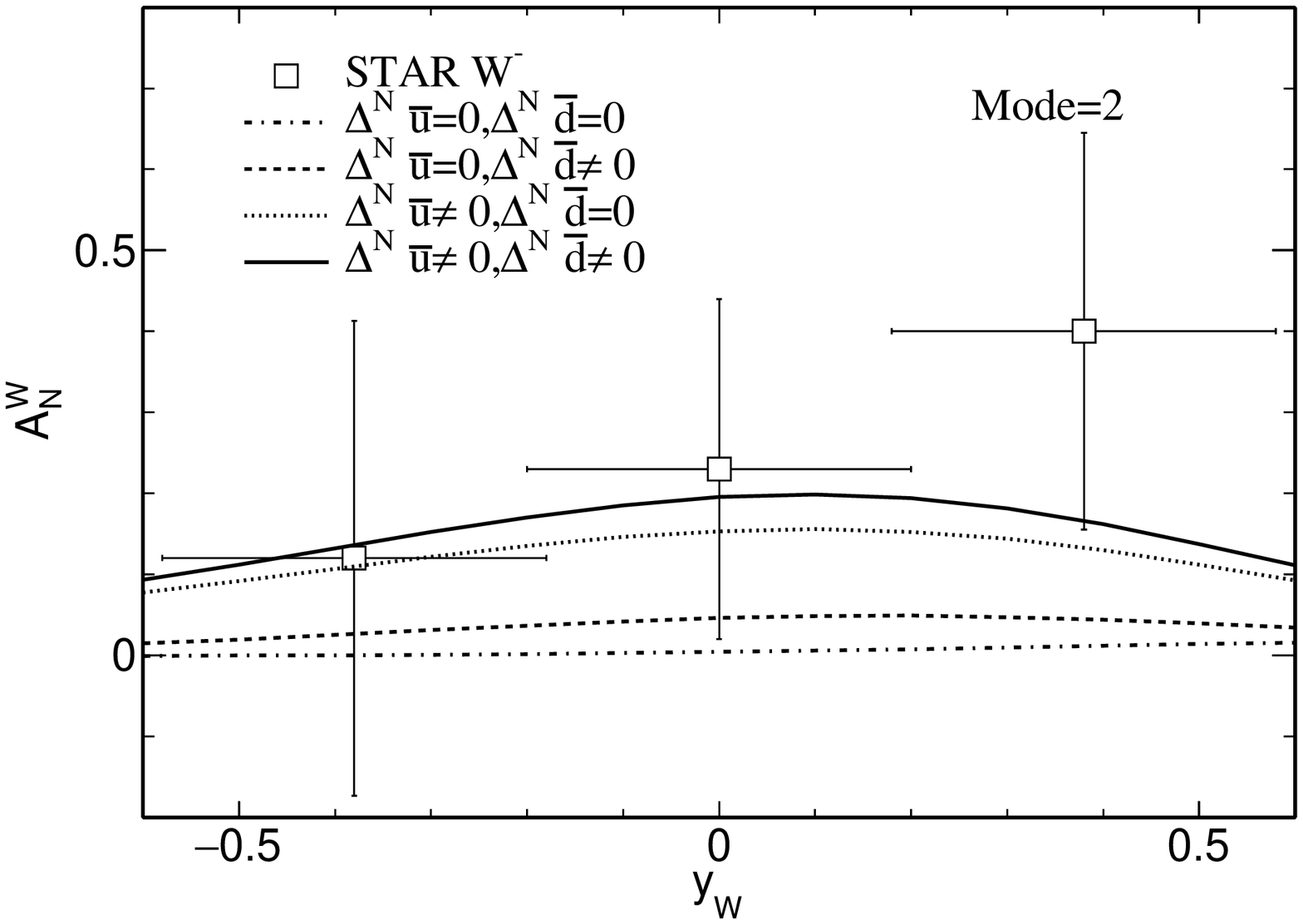}}		\subfigure[$A^{N}_{W^{+}}$.]{\includegraphics[width=0.40\textwidth]{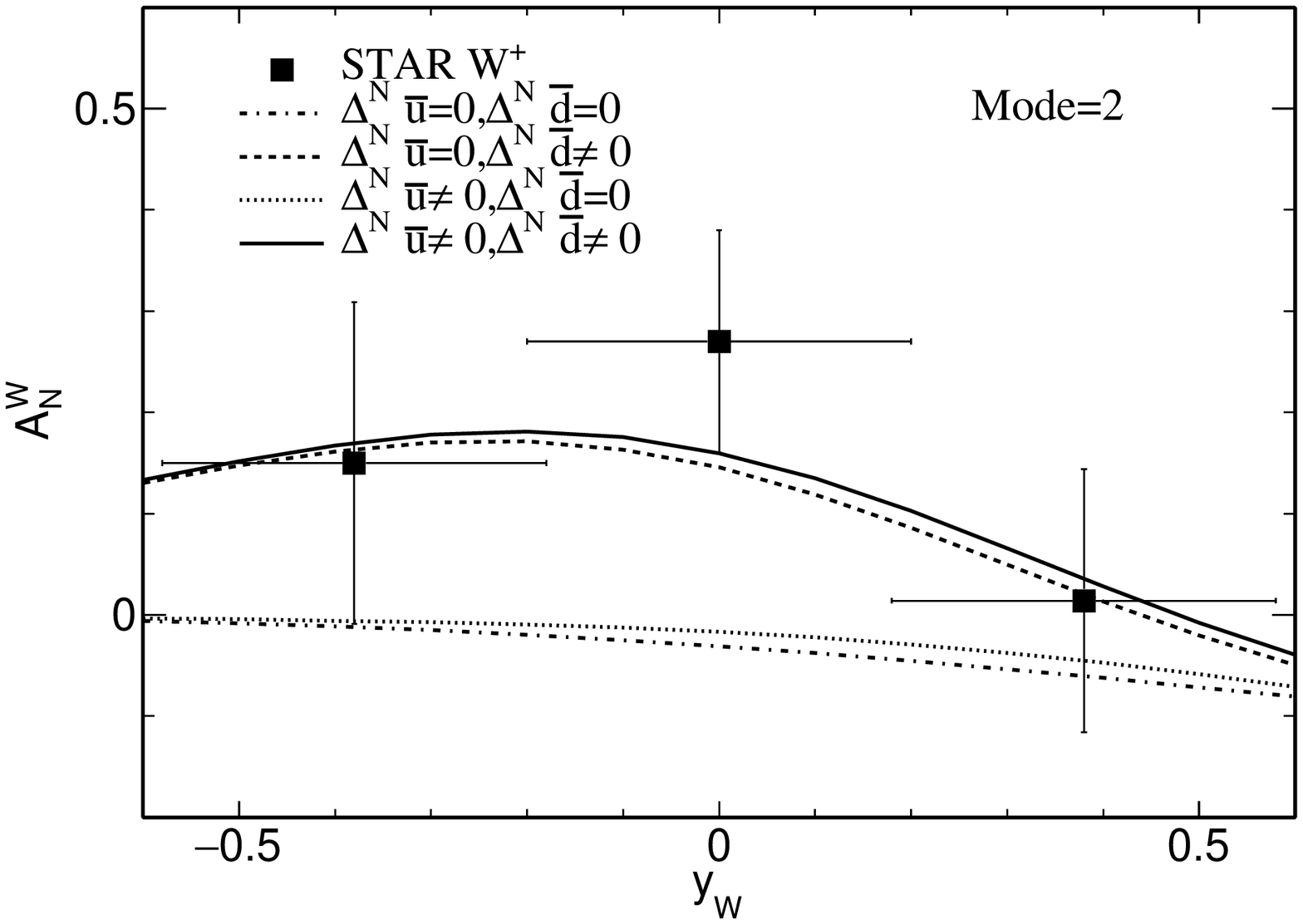}}
	\end{center}
	\vspace{-0.5cm}
	\caption{\label{fig:22} $A^{N}_{W^{\pm}}$ at $Q=M_W$ GeV for $\mathrm{Mode=2}$.}
\end{figure}

As for $W^{+}$, the contributions of $d$ sea Sivers functions are larger than that of $u$ sea Sivers functions, due to $\bar d\ll u$. Similarly, $u$ sea Sivers function plays an important role on $A^{W^{-}}_{N}$. In Ref.~\cite{Anselmino:2016uie}, the contributions of sea quark Sivers functions are quite smaller. In our work, we also consider the sea quark Sivers functions, just as the solid black curves shown in Fig.~{\ref{fig:21}} and Fig.~{\ref{fig:22}}. Besides, to exam the contributions from $u$ and $d$ sea quark Sivers functions, we calculate $A^{W^{\pm}}_{N}$ by setting one of the sea quark Sivers functions $\Delta^{N}\bar{q}=0$, e.g., the black dashed curves represent the contributions from $\Delta^{N}\bar{d}\ne 0$ with $\Delta^{N}\bar{u}=0$, while the black dotted curves stand for the contributions from $\Delta^{N}\bar{u}\ne 0$  with $\Delta^{N}\bar{d}=0$.

\begin{figure}[H]
	\begin{center}		
		\subfigure[$x f^{\perp(1)}_{1T}(x)$.]{\includegraphics[width=0.40\textwidth]{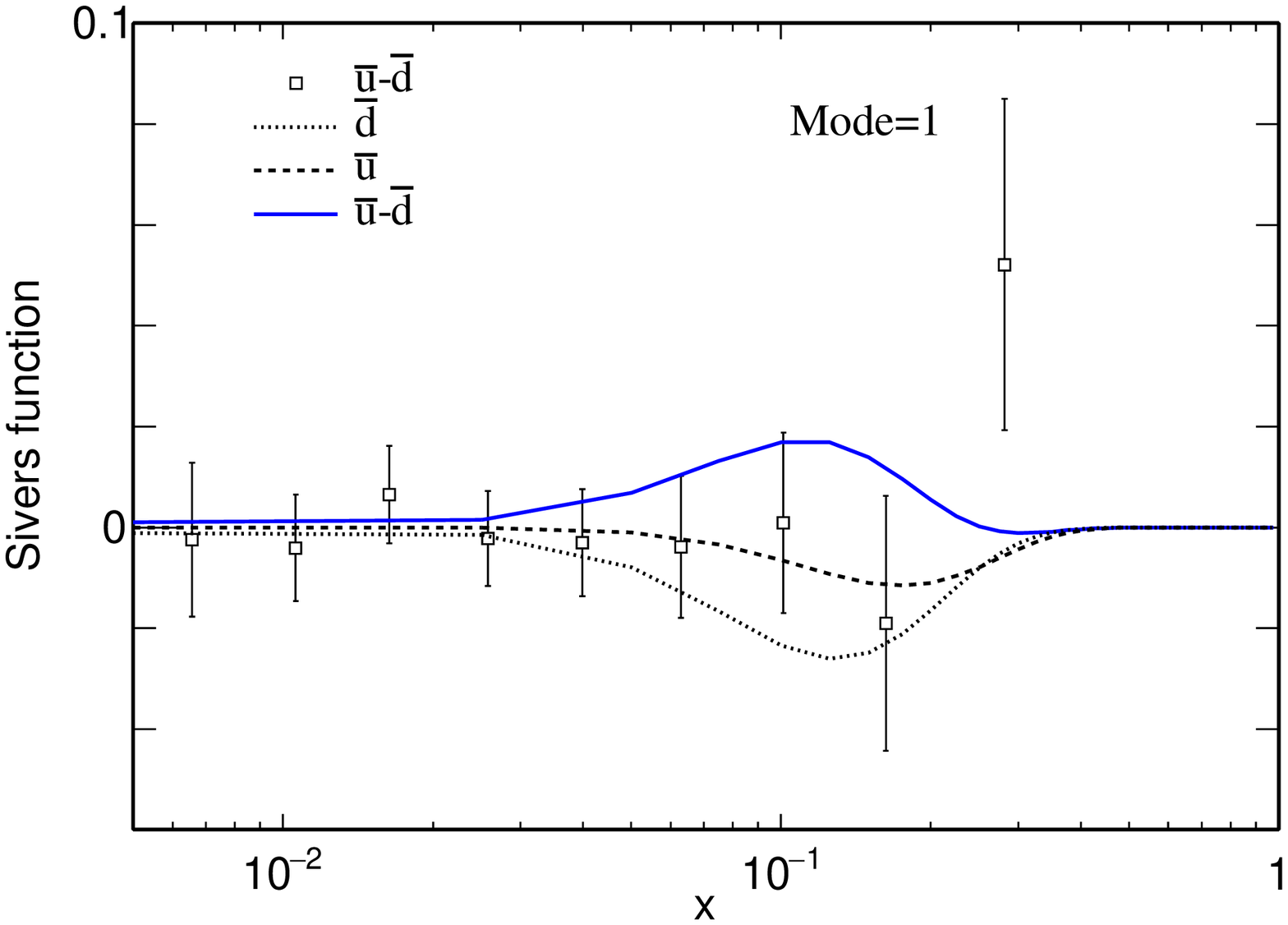}}
		\subfigure[$x f^{\perp(1)}_{1T}(x)$.]{\includegraphics[width=0.40\textwidth]{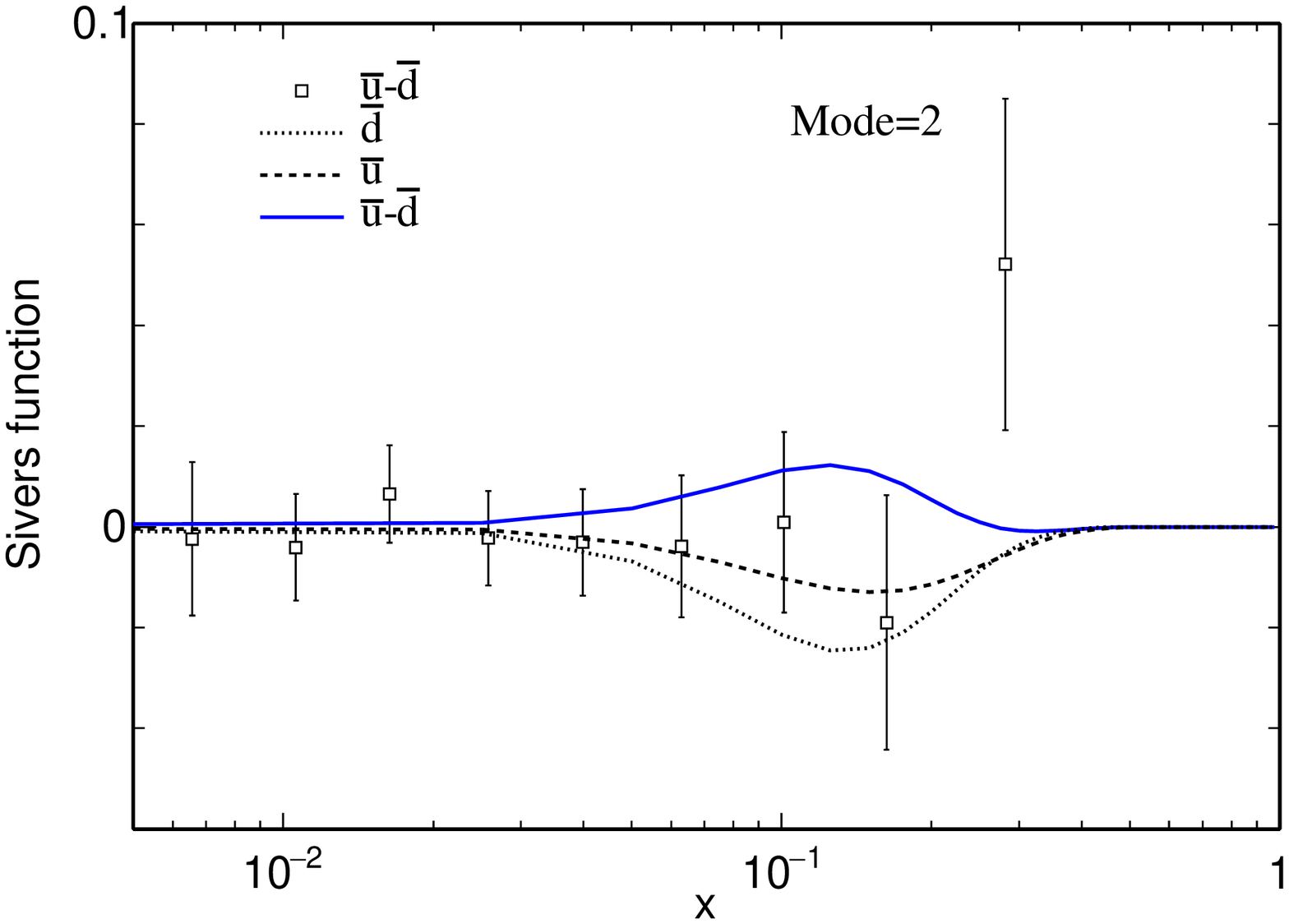}}
	\end{center}
	\vspace{-0.5cm}
	\caption{\label{fig:20} $f^{\perp(1)}_{1T}(x)$ at $Q=2$ GeV for $\mathrm{Mode=1,2}$. (b): $\bar{u}-\bar{d}$ denotes the extracted result from ~\cite{Martin:2017yms}.}
\end{figure}

\begin{figure}[H]
	\begin{center}		\subfigure[$Ratio^{\bar{u}}$.]{\includegraphics[width=0.40\textwidth]{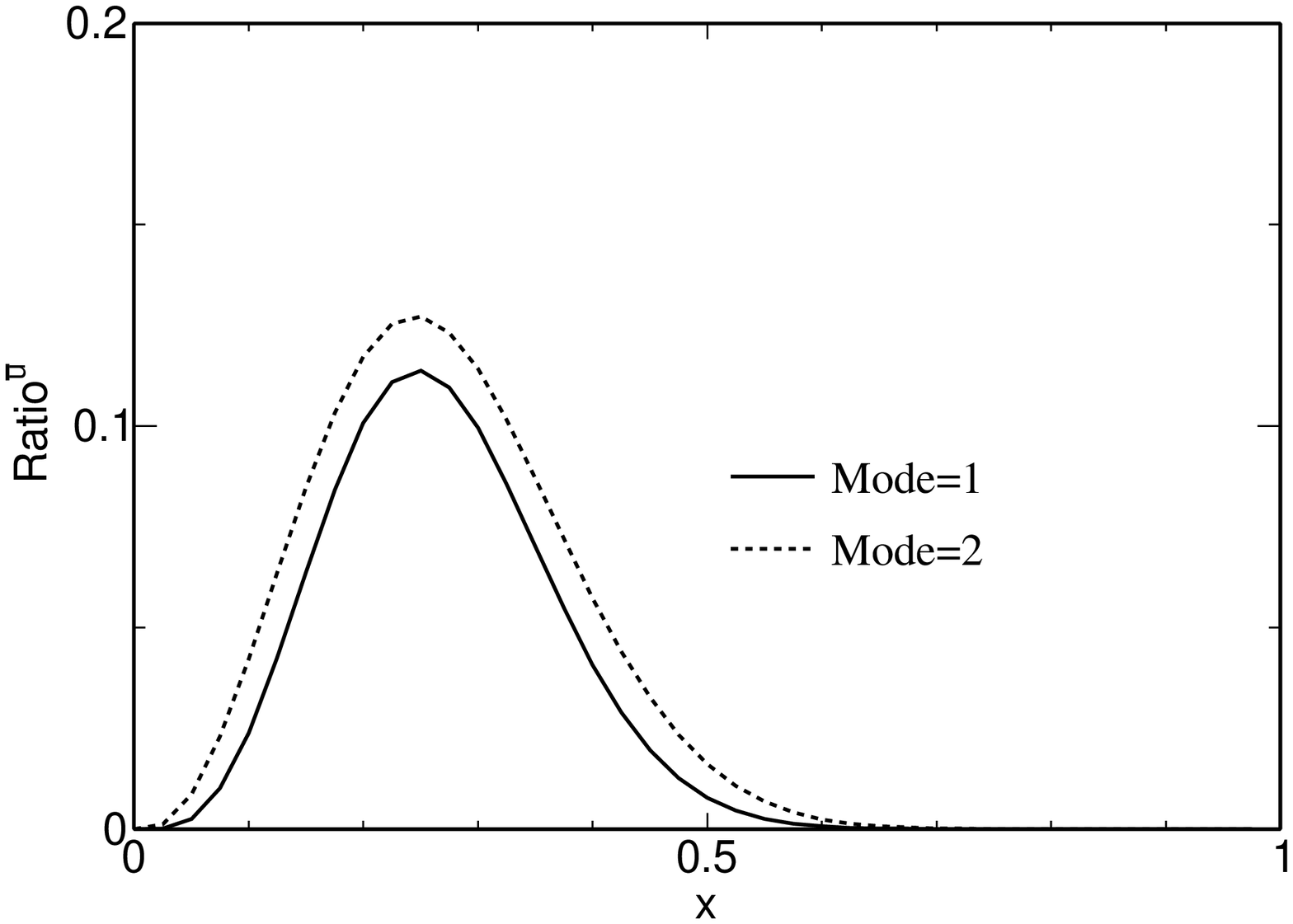}}
		\subfigure[$Ratio^{\bar{d}}$.]{\includegraphics[width=0.40\textwidth]{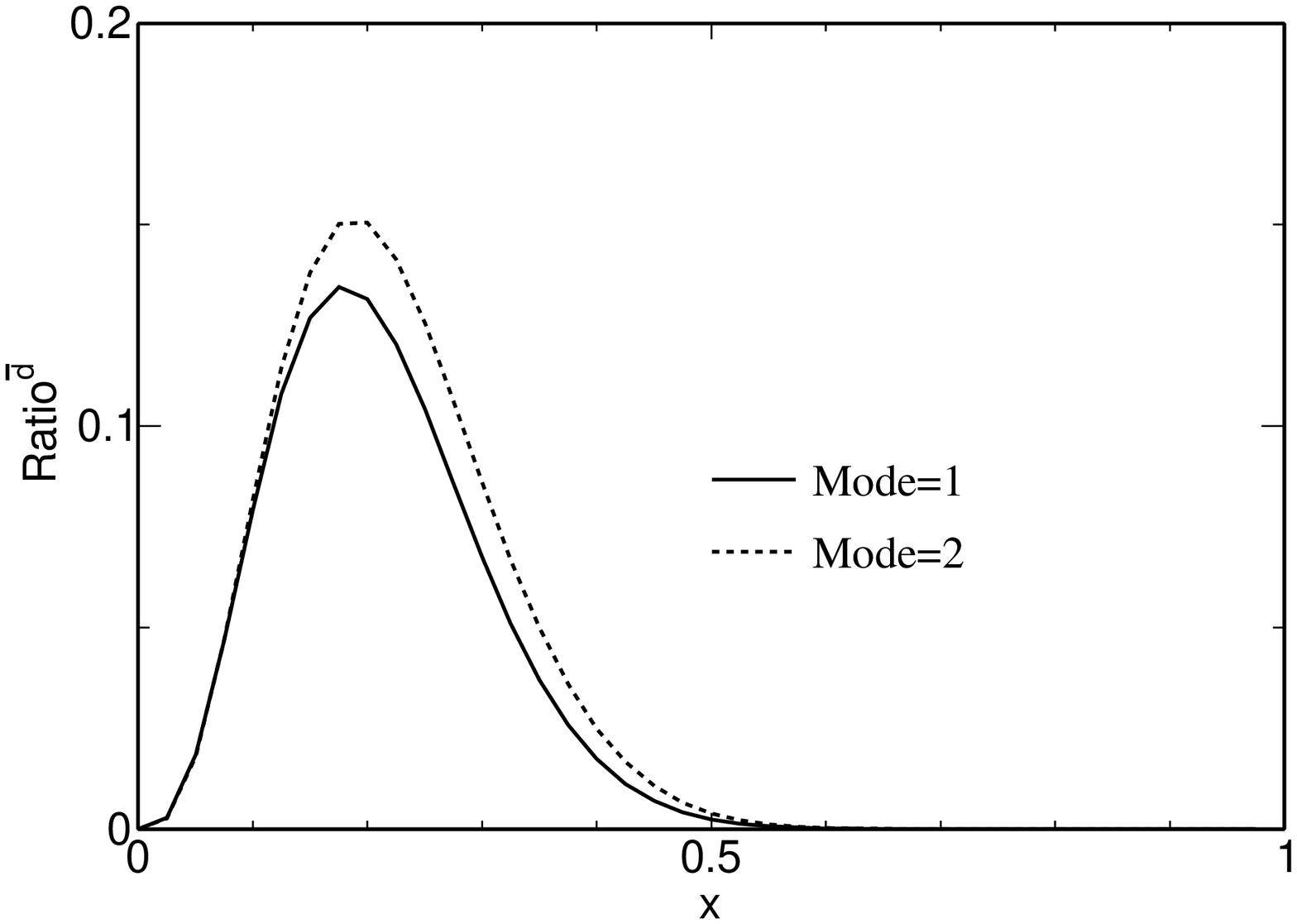}}
	\end{center}
	\vspace{-0.5cm}
	\caption{\label{fig:24} $|\Delta^{N}q(x)|/2q(x)$ at $Q=M_W$ GeV for $\mathrm{Mode=1,2}$.}
\end{figure}
From Table~\ref{table1}, by comparing the two different modes, we observe that $u$ and $d$ sea Sivers functions have the same sign, and their signs agree with that of $d$ valence quarks and are opposite to that of $u$ valence quarks, for better descriptions of the data. In Fig.~{\ref{fig:21}} and Fig.~{\ref{fig:22}}, the calculated $A^{W^{\pm}}_{N}$ can match the data with sizable sea quark Sivers functions. Besides, a good description of the shape of $A^{W^{-}}_{N}$ depends on negative valued $u$ sea quark Sivers functions mainly, while the good reproduction of the shape of $A^{W^{+}}_{N}$ depends on negative valued $d$ sea quark Sivers functions
mainly. The results of sea quark Sivers functions can be comparable with the extracted results of the isotriplet $\bar{u}-\bar{d}$ component from Ref.~\cite{Martin:2017yms} in Fig.~{\ref{fig:20}}, where we notice similar results for the two modes of sea Sivers functions.  In all modes, the ratios of ${\Delta^{N}\bar q(x)}/{\bar q(x)}$ with $q=u~\mathrm{or}~d$ satisfy the general relation Eq.~\ref{eq:pos} as shown in Fig.~\ref{fig:24}.

\begin{figure}[H]
	\begin{center}
		\subfigure[$A^{N}_{Z^{0}}$.]{\includegraphics[width=0.40\textwidth]{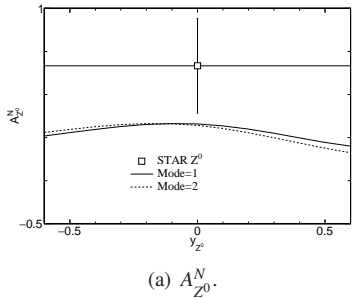}}
	\end{center}
	\vspace{-0.5cm}
	\caption{\label{fig:25} $A^{N}_{Z^0}$ at $Q=M_Z$ GeV for $\mathrm{Mode=1,2}$.}
\end{figure}

Based on the obtained sea quark Sivers functions in Table~\ref{table1}, we can also calculate $A^{Z^{0}}_{N}$ of $Z^{0}$ bosons as shown in Fig.~{\ref{fig:25}}.
From Fig.~{\ref{fig:25}}, we can see that the results seem to be smaller than the detected data, though our results are actually larger when comparing with those in Ref.~\cite{Anselmino:2016uie}.

From above discussions, we know that the theoretical calculations of $A^{W^{\pm}}_{N}$ could match the experimental data with sizable sea quark Sivers functions. Besides, the $x$-dependent relation of sea Sivers function could provide better description of the shapes of $A^{W^{\pm}}_{N}$ in comparison with those predicted from previous parameterizations and model calculations.

It is interesting to compare our results of $u$ and $d$ sea Sivers functions with the sea quark helicity distributions from longitudinally polarized pp collisions~\cite{Tian:2017xul}, in which the spins of sea quarks are also sizable, i.e., $\Delta\bar{u}>0$ and $\Delta\bar{d}<0$ with different signs. The sea Sivers functions of $u$ and $d$ quarks in transversely polarized nucleon are also sizable, but tend to have the same sign, i.e., $\Delta^N\bar{u}<0$ and $\Delta^N\bar{d}<0$. What is more,
based on statistical consideration~\cite{deFlorian:2009vb,Bourrely:2013qfa}, an intuitive explanation can be given to explain the signs of $u$ and $d$ sea quark helicity distributions,
while there still lacks a physical picture to understand the signs of $u$ and $d$ sea quark Sivers functions in the transverse polarized case.

\section{Summary}\label{sec:ww3}
In summary, we investigate the contributions from sea quark Sivers functions to the single-spin asymmetries $A^{W^{\pm}}_{N}$ of $W^{\pm}$ bosons in transversely polarized pp collisions. To confront with the experimental data at RHIC, we adopt two different modes of Sivers functions in our calculations. It is shown that $A^{W^{\pm}}_{N}$ are sensitive to the quark Sivers functions, especially the sea quark Sivers functions.
This study provides an intuitive picture about the role played by the single-spin asymmetries $A^{W^{\pm}}_{N}$ on our understanding of the nucleon spin structure. The results indicate rather sizable Sivers functions of $u$ and $d$ sea quarks, with both of them having opposite signs to that of valence $u$ Sivers functions. The combination with previous results
of $u$ and $d$ sea quark helicity distributions from longitudinally polarized pp collisions ~\cite{Tian:2017xul} can enrich our knowledge on the importance of sea quark contributions to the nucleon spin.
Therefore further theoretical and experimental studies are needed to explore the Sivers functions of both sea quarks and valence quarks of the nucleon in more details.

\section*{Acknowledgments}
We acknowledge helpful discussions with Zhun Lu. This work is partially supported by National Natural Science Foundation of China (Grant No.~11475006).

\end{document}